\documentstyle[preprint,aps,floats,psfig]{revtex}
\begin{document}
%\twocolumn[
\hsize\textwidth\columnwidth\hsize\csname@twocolumnfalse\endcsname

\draft

\title{Thermal expansion and Gr\"uneisen parameters of amorphous silicon: 
A realistic model calculation} 
\author{Jaroslav Fabian and Philip B. Allen}
\address{Department of Physics, State University of New York at Stony Brook, 
Stony Brook, New York 11794-3800}
\maketitle

\vspace{1cm}

\begin{abstract}

Using a realistic model, the mode Gr\"uneisen parameters $\gamma$
and the temperature dependent coefficient of linear thermal expansion
$\alpha(T)$, are calculated for amorphous silicon.
The resulting $\gamma$ values differ from the crystalline case
in having all diversity suppressed, except for a minority of
high-frequency localized and low-frequency resonant modes.
The latter have very large, mostly negative $\gamma$ (up to $-31$),
caused by volume driven internal strain. As a result, the
values for $\alpha(T)$ are lower than those of crystalline silicon
and are sample dependent.

\end{abstract}

\vspace{2em}
%]
%\narrowtext
%\tighten
\newpage

Unlike thermal conductivity and specific heat, 
which have ``universal'' features, thermal expansion is 
a special property of each glass. The coefficient of linear expansion 
$\alpha$ can be either positive or negative, with magnitude sensitive
to sample preparation methods \cite{barron}. This is true even at very low 
temperatures ($T\alt$1 K), where $\alpha$ is believed to 
be associated with tunneling modes \cite{galperin}
and large dispersion of $\gamma$ values is found \cite{negative}.

Here we present an analysis of $\alpha(T)$ for an atomistic model of
amorphous silicon, an important electronic material. The value of 
$ \alpha(T)$ has  been measured only at T=383 K \cite{witvrouw}, making
theoretical modelling particularly useful. Our calculation
shows that the value of $\alpha(T)$ is lower than that of 
crystalline silicon.
Deviations from the crystalline values are more dramatic at lower
temperatures, and are caused by volume driven internal strain,
which makes the thermal expansion sample
 dependent.
The predicted Gr\"uneisen parameters display a surprising 
simplicity, which we interpret as evidence for special, 
possibly generic, properties of vibrations in glasses.
Specifically, we find that the majority of modes which
are neither localized nor ballistically propagating
(we call them ``diffusons''), have a property of
{\it global indistinguishability}, whereas high-frequency
localized and low-frequency resonant modes are distinguished
by the special structural imperfections at which they
have largest amplitudes. In particular, the resonant modes 
have a considerable  dispersion of $\gamma$ values (from -31 to 4),
consistent with our finding that internal strain
is largest at the centers of these modes. 
Below 1K,  very large average magnitudes of $\gamma$ ($|\gamma|\agt 10$)
were measured  in many glasses \cite{negative} (``normal'' values are 
about one). 
Although the size of our model does not allow us to go as low as
1K, we show that the resonant modes provide a likely origin of this 
anomaly. Finally, we establish a close link between the mode 
Gr\"uneisen parameters and the mode  bond-stretching character.

Insofar as a complete set of normal modes can be defined
(i.e., the modes decay on timescales much larger than their periods),
one expects vibrational entropy $S$ to be well approximated by  
%\begin{eqnarray} \label{eqn:entropy}
$
S=k_B\sum_a[(n_a+1)\log(n_a+1)-n_a\log n_a],
$
%\end{eqnarray}
where $n_a=(\exp(\hbar\omega_a/k_BT)-1)^{-1}$ is the average
equilibrium occupation number of the mode of frequency $\omega_a$,
$a$ being a counting label going from 1 to $3N$. $N$ is the number of
atoms and $k_B$ is the Boltzmann constant. The frequency $\omega_a(V,T)$
may depend on both volume $V$ and temperature.
Using the standard thermodynamic relation 
$3\alpha=\kappa_{\rm T}(\partial S/\partial V)_T$, with 
isothermal compressibility $\kappa_{\rm T}$,  we get (see, e.g., Ref.
\cite{cowley})
\begin{eqnarray}\label{eqn:expansion}
\alpha(T)
=\frac{\kappa_{\rm T}}{3V}\sum_{a}c_a\gamma_a.
\end{eqnarray}
%\begin{eqnarray}\label{eqn:gruneisen}
%\gamma_a=-\partial \log \omega_a/\partial \log V.
%\end{eqnarray}
Here $c_a=k_B (\hbar\omega_a/k_BT)^2 n_a(n_a+1)$ is the specific heat
of a harmonic oscillator, and $\gamma_a$, given by
$\gamma_a=-\partial \log \omega_a/\partial \log V$,
 are known as the ``mode Gr\"uneisen
parameters''; they measure how sensitive the vibrational eigenfrequencies 
are to the change of volume. Their knowledge is essential not only for 
$\alpha(T)$, but also for the  interpretation of  the internal friction and 
sound attenuation experiments \cite{sound,jaro}.

Perturbation theory (see, e.g., Ref. \cite{klein}) gives the following 
formula for $\gamma_a$,
\begin{eqnarray} \label{eqn:grun}
\gamma_a&=&-\frac{1}{6\omega^2_a}
\sum_{ijk}\sum_{\alpha\beta\gamma}
\Phi^{ijk}_{\alpha\beta\gamma}\frac{e^{a*}_{i\alpha}
e^a_{j\beta}}{\sqrt{M_iM_j}}(r_{k\gamma}+\bar{r}_{k\gamma}), \\
\label{eqn:internal}
\bar{r}_{i\alpha}&=&-\sum_a\frac{1}{\omega_a^2}\sum_{jk}\sum_{\beta\gamma}
 \Phi^{jk}_{\beta\gamma}\frac{e^{a*}_{i\alpha}e^a_{j\beta}}{
\sqrt{M_iM_j}} r_{k\gamma}.
\end{eqnarray}
Here $\Phi^{ij}_{\alpha\beta}$ ($\Phi^{ijk}_{\alpha\beta\gamma}$) are
the coefficients of the quadratic (cubic) terms of the Taylor expansion of 
potential energy in terms of the displacements $u_{i\alpha}$ of atoms $i$ from 
equilibrium in the direction $\alpha=(x,y,z)$, 
and $r_{i\alpha}$ is the position vector of the $i$th atom.
The vibrational eigenvectors 
$e^a_{i\alpha}$ are normalized ($\sum_{i\alpha} |e^a_{i\alpha}|^2$=1),
and $M_i$ is the mass of atom $i$.
The sums in Eq. \ref{eqn:grun} and \ref{eqn:internal} are over all atoms.
Because of periodicity, in a crystal the label $a$ can be written as
({\bf Q},$\lambda$), denoting wave vector and polarization.
The eigenvectors $e_{i\alpha}({\bf Q},\lambda)$ are  proportional
to $e^{i{\bf Q}\cdot{\bf r}_i}/\sqrt{N}$, and the resulting 
crystalline phase coherence allows a simplification of the sum in 
Eq. \ref{eqn:grun}  to the neighborhood of a single small
unit cell. By contrast, normal modes in a glass have no 
{\it a priori} quantum numbers, and quantitative insight is best achieved
by numerical diagonalization for finite models.

If a solid is subject to an infinitesimal homogeneous isotropic strain
$\epsilon$, its atomic coordinates change to $r'_{i\alpha}=
r_{i\alpha}(1+\epsilon)$, and volume to $V'=V(1+3\epsilon)$. 
Unless atoms in the solid are fixed by symmetry, $r'_{i\alpha}$ are 
not the equilibrium coordinates of the solid with volume $V'$.  
This happens in glasses. After the strain is applied, the atoms
relax to new equilibrium positions $r'_{i\alpha}+\epsilon\bar{r}_{i\alpha}$.
This is the origin of the {\it internal strain} (Eq. \ref{eqn:internal}) 
occuring during the 
thermal expansion of glasses (and crystals with atoms not
in centers of symmetry \cite{barron}). 
In silicon crystal $\bar{r}_{i\alpha}\equiv 0$.

The interactions between silicon atoms are represented by the 
Stillinger-Weber (SW) potential \cite{stillinger}, which performs
well when applied to elastic and vibrational properties of different
silicon phases \cite{broughton,allen2,fabian}. 
For the atomic coordinates of amorphous silicon
we take the model introduced and described in detail in  Ref.
\cite{allen2}. The algorithm by Wooten, Winer, and Weaire (WWW)
\cite{wooten} creates a random network structure of silicon
atoms, which are further relaxed to a local minimum of the SW potential.
Here we use 1000 atoms arranged in a cube of side 27.543 \AA, with
periodic boundary conditions. The energy/atom of this model
is $-4.102$ eV, which is  $\approx 5\%$ higher than the 
energy/atom of crystalline silicon
with SW potential. Also, the model agrees very well with
the neutron structure factor $S(Q)$ measured by Kugler 
{\it et al}. \cite{kugler}.

To calculate $\alpha(T)$ we need the values for  $\kappa_{\rm T}$.
The corresponding perturbation
formulas for $\kappa_{\rm T}$, which take into account internal 
strain, can be found in Ref. \cite{maradudin}.
We obtain the value 
$1.13\times 10^{-12}$ $\rm cm^2/dyn$ for the present model of amorphous
silicon. For crystalline silicon, the SW potential gives the value of 
$0.986\times 10^{-12}$ \cite{cowley1}, agreeing well with
the experimental value of $1.02 \times 10^{-12}$ \cite{cowley1}. 
Also, for silicon crystal, we must sum over
the {\bf Q} points, which we do  using the tetrahedron 
method with 1772 tetrahedra in the irreducible wedge of the Brillouin zone.
The results are shown in Fig. \ref{fig:1}, together
with the measured data.

Compared with experiment \cite{ibach}, $\alpha(T)$ of silicon crystal
is reproduced very well at $T\agt$200 K. 
At lower $T$ the SW potential does not reproduce the observed
negative values of $\alpha$. Negative $\alpha$ has been
succesfully explained \cite{white} by negative values of
$\gamma$ for the low energy transverse acoustic (TA) branch.
As shown in Fig. \ref{fig:2}(A), our $\gamma$ values are too
weakly negative in this regime. In Fig. \ref{fig:1} it is
predicted that $\alpha(T)$ for amorphous silicon is somewhat lower 
than $\alpha(T)$ for the crystal. 
For comparison, Fig. \ref{fig:1} also shows $\alpha(T)$ 
without considering internal strain ($\bar{r}_{i\alpha}$ in Eq. 
\ref{eqn:grun} is set to zero), and the calculation based on
a 216 atom model of amorphous silicon. Internal strain
reduces the values of $\alpha$ by almost 30\% at high
$T$. At lower $T$ the values become negative. 
The model dependence is also clear.
The 216 atom model is based on the same WWW algorithm and SW 
potentials, but is more topologically constrained (e.g., no 
four-fold rings are allowed), and has lower energy/atom (by 
$\approx 0.5\%$) and higher density (by $\approx 3\%$) than the 
1000 atom model. Its $\alpha(T)$ is higher and closer to the crystalline
case and the measured value. 
This is not surprising. Measurements on silica \cite{barron,white1} 
showed that annealing history (or density) markedly changed its thermal
expansion at all temperatures. For example, pure $\rm SiO_2$
aged at $\rm 1400^\circ C$ and quenched, has $\alpha(T)$ at high $T$
lower by up to $50\%$ than that slowly aged at $\rm 1000^\circ C$.
Our calculation predicts that future experiments should see similar
behavior in amorphous silicon as well. 

To understand the behavior of $\alpha(T)$ one has to look at
the frequency dependence of Gr\"uneisen parameters. 
The values of $\gamma$ for silicon crystal, for more than 
1000 randomly chosen {\bf Q} points from the irreducible wedge 
of the Brillouin zone, are in Fig. \ref{fig:2}(A).  
Since there is a degenerate surface in {\bf Q} space for a given
$\omega$, $\gamma({\bf Q},\lambda)$ has a distribution of values
at each $\omega$. These values are further split according
to the branch $\lambda$ of corresponding phonons.
Particularly striking polarization effects appear at low $\omega$,
where TA phonons form a distinct broad band
of $\gamma_a$. 

It is instructive to compare Fig. \ref{fig:2}(A)
with Fig. \ref{fig:2}(B), where we plot the mode bond-stretching 
parameter $S_a$,
\begin{eqnarray}
S_a=\left[\frac{\sum_{<i,j>}\left|({\bf u}^a_i-
{\bf u}^a_j)\cdot {\bf n}_{ij}\right|^2}
{\sum_{<i,j>}\left | {\bf u}^a_i-{\bf u}^a_j \right|^2}\right]^
{\frac{1}{2}},
\end{eqnarray}
introduced in Ref. \cite{zotov} (we modified slightly the original formula
which did not yield the same $S_a$ for different but degenerate $a$).
The summation is over all atom pairs $<i,j>$, where $j$ is within
2.8 \AA  (the distance where the first nearest neighbor peak in the
pair correlation function of amorphous silicon ends \cite{kugler}) from $i$,
and ${\bf n}_{ij}$ is the unit vector in the direction of the bond $<i,j>$.
When $S_a$ is close to one, mode $a$ is mostly
bond-stretching, while values closer to zero indicate bond-bending
modes. The similarity between the two figures is striking. Except at very
low $\omega$, one is tempted to write ``$\gamma_a \approx {\rm const}\times 
S_a$.'' The reason for such close relation is that anharmonicity,
reflected in $\gamma_a$, is much greater in the bond-stretching
part of the interatomic interaction than in the bond-bending part.

We now discuss the behavior of $\gamma_a$ in  amorphous silicon,
whose vibrational states were discussed earlier \cite{allen2,fabian}.  
At low frequencies ($\alt 15$ meV) the modes are {\it propagons},
acoustic-phonon-like vibrations propagating ballistically for distances
greater than their wavelength. 
Some of the propagons are resonantly trapped at certain places in the
sample, with reduced amplitude elsewhere (but not exponential decay as
in a localized state); but at high $\omega$ there is no
continuum of extended states to decay into, and vibrational 
amplitudes decay exponentially with distance.
Properties of these
resonant modes were studied in different glassy systems \cite{resonant},
and for our model will be reported elsewhere \cite{resonant1}.  
Modes between 15 meV and 71 meV are {\it diffusons}, extended 
states carrying energy diffusively. Since they form the majority 
of the spectrum, diffusons dominate the thermal properties
of amorphous silicon at temperatures from several kelvins up 
to the melting point. Above $\omega_c\approx 71$ meV, the mobility edge,
the modes are {\it locons}, localized states.

The calculated  $\gamma_a$ are shown in Fig. \ref{fig:3}(A).
Three regions can be clearly distinguished. 
(i) At low frequencies the values are very scattered, unusually large, 
and mostly negative.
(ii) In the region of diffusons, after a monotonous increase, 
$\gamma_a$ becomes almost constant at $\approx$30 meV, and its values
collapse into a very narrow region. 
(iii) At locon frequencies $\gamma_a$ spreads out again, but much less 
than at low $\omega$.

Why are the low frequency values of $\gamma$ so large? At most one would 
expect the data to be scattered between 0 and 1 due to a partial presence
of {\bf Q} and $\lambda$ for propagons. We find that the cause of this
anomalous behavior is the resonant modes. As already shown \cite{resonant},
these modes have very large amplitudes at groups of under-coordinated atoms.
Similar groups of over-coordinated atoms are responsible for locons.
Since under-coordination implies ``softness'',
it is  natural that the internal strain $r_{i\alpha}$ is also largest
at these sites. This is why the magnitudes of $\gamma$ are large for 
the resonant modes. The dispersion of values, also seen for locons,
is explained by different nature of topological defects where the modes
have largest amplitudes.

The bands seen in $\gamma({\bf Q},\lambda)$, Fig. \ref{fig:2}(A),
of crystalline silicon, are suppressed in the ``diffuson'' portion
of the spectrum of amorphous silicon. This has the following interpretation.
In a crystal, knowledge of the pattern of a normal mode in a few adjacent
unit cells allows one to predict the pattern in distant regions.
In a glass, such a prediction is not possible without a very
large  computer calculation. More surprising, different normal modes
in the ``diffuson'' part of the spectrum are not globally
distinguishable. If one knows the pattern of displacement of two modes
($a1$ and $a2$) of similar frequencies in one spatial region,
one could not recognize which pattern was $a1$ and which $a2$ in a
distant region. There seem not to be properties other than $\omega_a$
which can classify these modes.
 
Finally, Fig. \ref{fig:3}(B) shows $S_a$ for amorphous silicon. 
Note the surprising uniformity of bond-stretching character (see also
Ref. \cite{zotov}), even in the region of propagons where memory
of polarization character might have been expected.
As in the
crystalline case, $S_a$ closely follows  $\gamma_a$, except
at low frequencies where the values of $\gamma$ are much more influenced
by dynamics than by vibrational pattern.

We thank J. L. Feldman and S. Bickham for stimulating                 
discussions.  This work was supported by NSF Grant No. DMR 9417755.

\newpage

\begin{figure}
%\centerline{\psfig{file=fig1.ps,width=1.\linewidth,angle=0}}
\caption{Calculated and measured linear thermal expansion $\rm \alpha(T)$
for crystalline and amorphous silicon. IS: internal strain.
}
\label{fig:1}
\end{figure}

\begin{figure}
%\centerline{\psfig{file=fig2b.ps,width=1\linewidth,angle=0}}
\caption{Calculated Gr\"uneisen parameters $\gamma_a$ (A) and bond
stretching parameters $S_{a}$ (B) for crystalline silicon
as a function of frequency. In (A) the polarization labels
are: TA - transverse acoustic, LA - longitudinal acoustic,
LO - longitudinal optical, TO - transverse optical.
The solid line is the vibrational densities of states (DOS)
in arbitrary units.}
\label{fig:2}
\end{figure}

\begin{figure}
%\centerline{\psfig{file=fig3.ps,width=1\linewidth,height=1\linewidth,angle=0}}
\caption{
Calculated Gr\"uneisen parameters $\gamma_a$ (A) and
bond-stretching character $S_{a}$ (B) of amorphous silicon as a function
of frequency. The vertical line at $\omega_c\approx 71$ meV is the
mobility edge. The scale in (A) is split by the horizontal line
at $-1$ to emphasize the large negative values of $\gamma$ at
low $\omega$. The solid line is DOS in arbitrary units.}
\label{fig:3}
\end{figure}

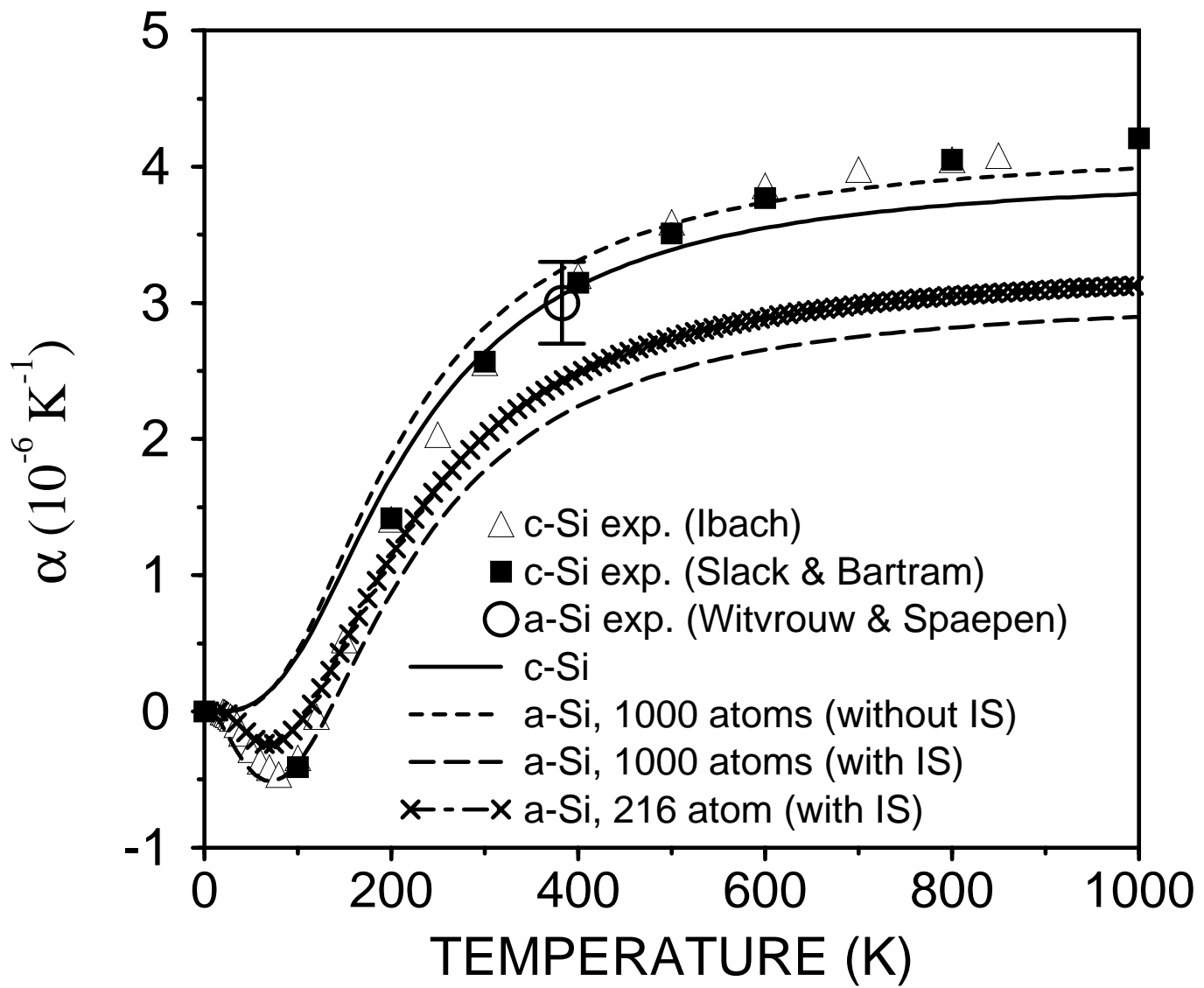
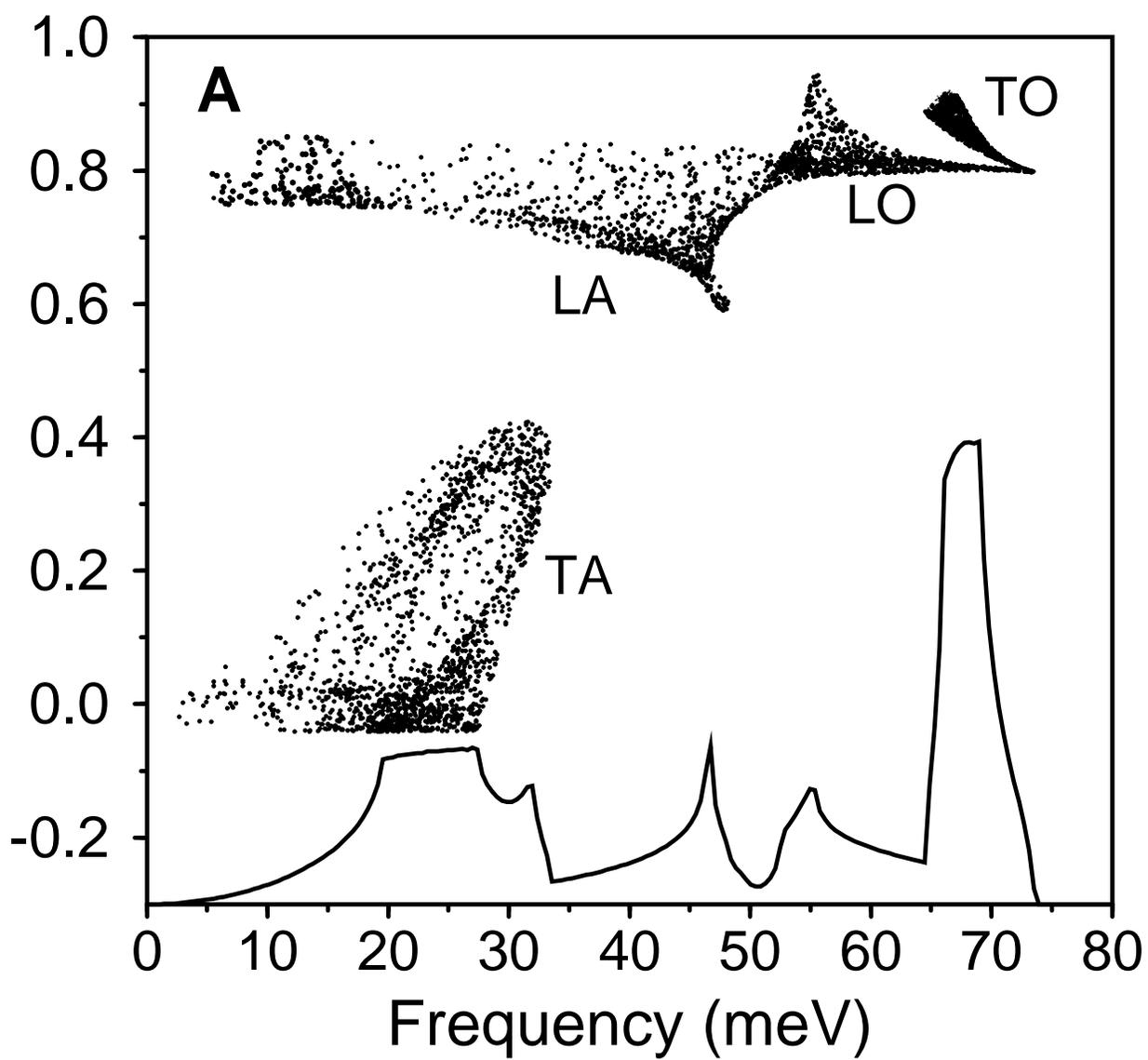
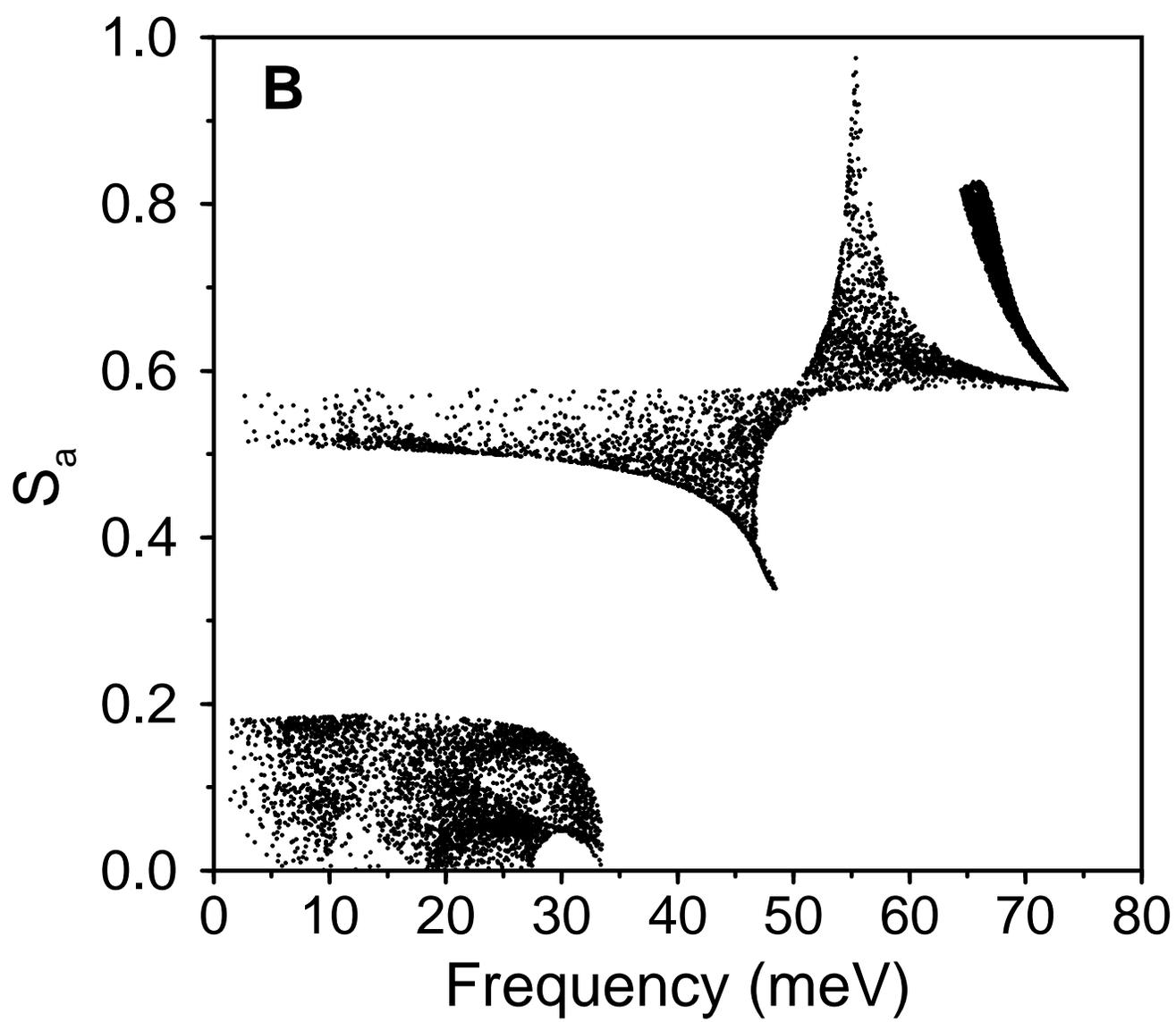
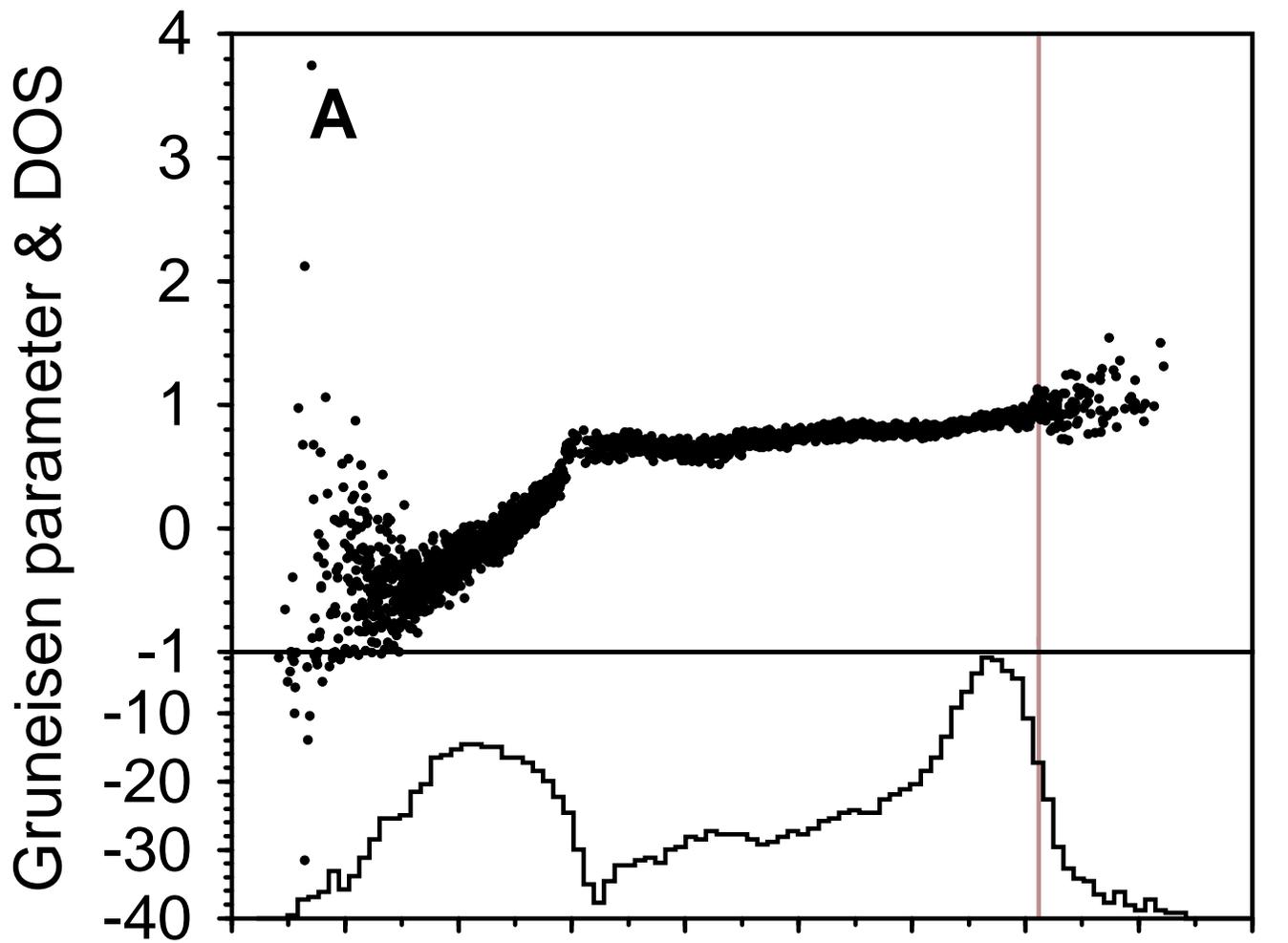
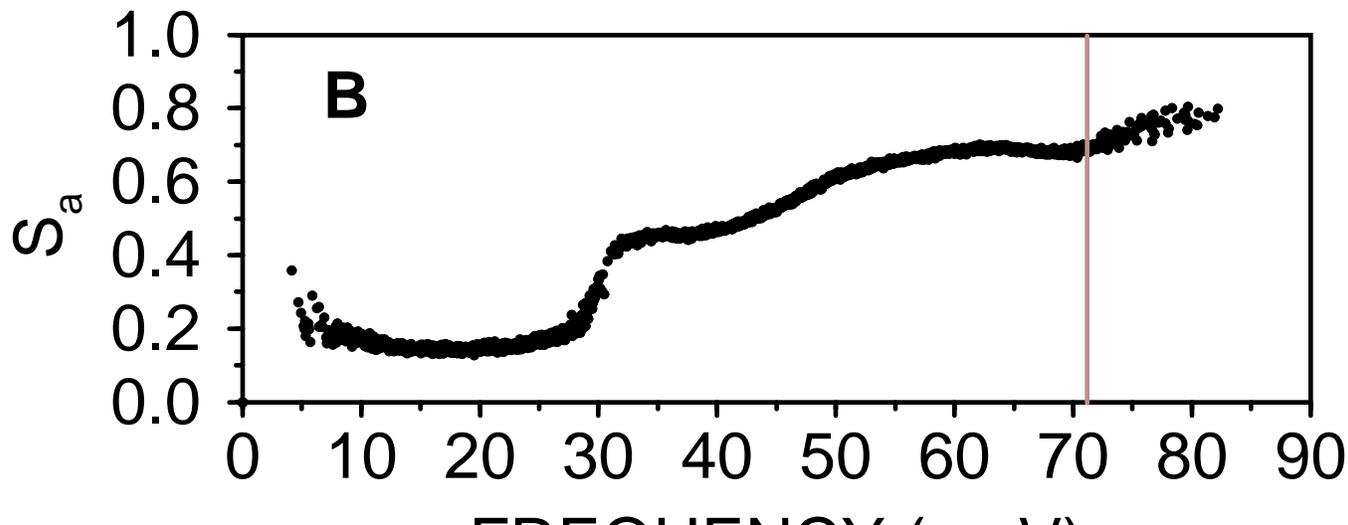
\end{document}